\documentclass[twocolumn,showpacs,superscriptaddress,nofootinbib,floatfix]{revtex4}
\usepackage{graphicx}

\begin{document}
% A useful Journal macro
\def\Journal#1#2#3#4{{#1} {\bf #2}, #3 (#4)}
\title{\bf The Centrality Dependence of the Parton Bubble Model for high energy
heavy ion collisions and fireball surface substructure at RHIC}
\affiliation{City College of New York, New York City, New York 10031}
\affiliation{Brookhaven National Laboratory, Upton, New York 11973}
\author{S.J.~Lindenbaum}\affiliation{City College of New York, New York City, New York 10031}\affiliation{Brookhaven National Laboratory, Upton, New York 11973}
\author{R.S.~Longacre}\affiliation{Brookhaven National Laboratory, Upton, New York 11973}
\date{\today}% It is always \today, today,
 
\begin{abstract}
In an earlier paper we developed a QCD inspired theoretical parton bubble model
(PBM) for RHIC/LHC. The motivation for the PBM was to develop a model which 
would reasonably quantitatively agree with the strong charged particle pair
correlations observed by the STAR collaboration at RHIC in Au + Au central 
collisions at $\sqrt{s_{NN}}$ = 200 GeV in the transverse momentum range
0.8 GeV/c to 2.0 GeV/c. The model was constructed to also agree with the 
Hanbury Brown and Twiss (HBT) observed small final state source size 
$\sim$2fm radii in the transverse momentum range above 0.8 GeV/c. The model 
assumed a substructure of a ring of localized adjoining $\sim$2fm radius 
bubbles perpendicular to the collider beam direction, centered on the beam, at 
mid-rapidity. The bubble ring was assumed to be located on the expanding
fireball surface of the Au + Au collision. These bubbles consist of almost 
entirely of gluons and form gluonic hot spots on the fireball surface.
We achieved a reasonable quantitative agreement with the results of both
of the physically significant Charge Independent (CI) and Charge Dependent
(CD) correlations that were observed. In this paper we extend the model (PBME) 
to include the changing development of bubbles with centrality from the most 
central region where bubbles are very important to the most peripheral 
where the bubbles are gone. Energy density is found to be related to 
bubble formation and as centrality decreases the maximum energy density 
and bubbles shift from symmetry around the beam axis to the reaction plane
region causing a strong correlation of bubble formation with elliptic flow.
We find reasonably quantitative agreement (within a few percent of the total 
correlations) with a new precision RHIC experiment which extended the 
centrality region investigated to the range 0-80\% (most central to most 
peripheral). The characteristics and behavior of the bubbles imply
they represent a significant substructure formed on the surface of the
fireball at kinetic freezeout.
\end{abstract}
                                                                       
 
\maketitle

\section{Introduction and Review of Model} 

In the early eighties L. van Hove\cite{VanHove} proposed a bubble model as 
a way of finding convincing evidence for a Quark-Gluon Plasma (QGP). His model
based on a string calculation predicted that in one to perhaps a few 
observable localized rapidity bumps would appear in the final 
state of some events and he gave a prescription for experimentally finding
them. We searched for these in all relevant data, but we nor anyone else 
ever found experimental evidence for the rapidity bumps.

There were numerous other bubble models and some charge correlation models
proposed. Some examples of these are 
references\cite{Lopez,Pratt,Gyulassy,Dumitru,Drescher,LL}.
However, to our knowledge no significant experimental evidence has been 
found for any of these.

    In a previous publication\cite{PBM} we proposed a parton bubble model (PBM)
for central (impact parameter near zero) high energy heavy ion collisions at
RHIC/LHC which contains a substructure consisting of an 8fm radius single ring 
of a dozen adjoining 2f radius bubbles transverse to the collider beam 
direction, centered on the beam, and located at or near mid-rapidity on the
expanding surface of the fireball at kinetic freezeout. The bubble radius, 
and the bubble ring radius were estimated considering the Hanbury-Brown and 
Twiss (HBT\cite{HBT}) observations, and other general considerations utilizing 
the blast wave model. We assumed these bubbles (gluonic hot spots) are likely 
the final state result of quark-gluon plasma (QGP) formation. Thus this is the 
geometry for the final state kinetic freezeout of the QGP bubbles on the 
surface of the expanding fireball treated in a blast wave model. In the central
(near impact parameter 0) mid-rapidity region at RHIC we are observing the 
region where the highest energy densities and temperatures (parton energies) 
are produced. The experimentally observed $\sqrt{s_{NN}}$ = 200 GeV central Au 
+ Au collisions at RHIC\cite{centralproduction} produce initial energy 
densities\cite{adams} which exceed those predicted by lattice quantum 
chromodynamics (QCD) as sufficient for production of a QGP\cite{karsch}.
 
This single bubble ring resides at mid-rapidity on the surface of the 
expanding fireball at kinetic freezeout. Thus each bubble would emit a 
considerable fraction of final state particles (observed experimentally) 
resulting from the QGP state. The fraction of all the final state particles
from bubbles is $\sim$ 1/2. There would be very little re-interaction for 
particles emitted outward from the surface because the final state surface of 
the fireball is at kinetic freezeout. The bubble substructure (surrounded by 
cooler background) results in the lumpy surface of the fireball at kinetic 
freezeout. Section II presents more detail on the assumptions made, the 
development, and construction of the PBM.

The PBM successfully explained in a reasonably quantitative manner all of 
the particle pairs correlations in a precision STAR central production 
experiment\cite{centralproduction}. See Section 4 of Ref.\cite{PBM}
for this comparison. In Ref.\cite{PBM} some aspects of quark-quark 
recombination were compared with the PBM and a good agreement was obtained 
(see Section 5 PBM\cite{PBM}). 

In this paper we extend the model of Ref.\cite{PBM} to consider the case
of varying the centrality bins (Section III). We wish to compare with a new 
RHIC $\sqrt{s_{NN}}$ = 200 GeV minimum bias Au + Au analysis which covered the 
0-80\% centrality range\cite{centralitydependence}. 

This paper is organized as follows:

Section I is the Introduction. Section II summarizes the assumptions made
in the prior model (PBM) for the central region (0-10\% centrality\cite{PBM})
Section II also discusses the relevance and the reasoning behind these
assumptions. Section III discusses extending the PBM to the Parton Bubble
Model Extended (PBME) so that it becomes able to reasonably quantitatively 
fit and explain the new 0-80\% centrality high precision 
data\cite{centralitydependence}. Section IV discusses general characteristics
of the PBM. Section V presents and discusses a comparison of the 
experimentally determined Charge Independent (CI) correlation with PBME
as a function of \% centrality. Section VI presents and discusses a comparison
of the experimentally determined Charge Dependent (CD) correlation with PBME 
as a function of \% centrality. Section VII presents further details of the
PBME bubble correlation. Section VIII is the Summary and Discussion.

\section{Assumptions and Development of the PBM}

Our goal for the past 2 decades was to develop a model of bubble production
in relativistic heavy ion collisions assumed to originate from a QGP,
which could be reasonably quantitatively compared with relevant experimental
data of sufficient precision and scope. Thus hopefully we could obtain
convincing, or at least substantial evidence for the existence of bubble
production. Since this is a process which clearly involves strong 
non-perturbation QCD, pQCD calculations can serve only as a rough guide.

Thus we concluded we needed to obtain a strong hint from experimental data
as evidence for possible bubble substructure to proceed to build a realistic
model. The failures of obtaining significant experimental evidence for the 
many bubble models which did not incorporate such a strong hint from 
the experimental data led us to conclude that it was essential 
to obtain one from experimental observations. Then one could build 
the bubble model based on observations in order to provide quantitative
description of the data.

We utilize a two particle correlation function in the two dimensional (2-D)
space of $\Delta \phi$ vs $\Delta \eta$. The azimuthal angle $\phi$ of a
particle is defined by the angle of the particle with respect to the
vertical axis which is perpendicular to the beam axis and is measured in a
clock-wise direction about the beam. $\Delta \phi$ is the difference,
$\phi_1$ - $\phi_2$, of the $\phi$ angle of a pair of particles (1 and 2).
The pseudo-rapidity $\eta$ of a particle is measured along one of the beam
directions. $\Delta \eta$ is the difference, $\eta_1$ - $\eta_2$, of the
$\eta$ values of a pair of particles (1 and 2).
 
The two dimensional (2-D) total correlation function is defined as:
 
\begin{equation}
C(\Delta \phi,\Delta \eta)
=S(\Delta \phi,\Delta \eta)/M(\Delta \phi,\Delta \eta).
\end{equation}
                                                                            
Where S($\Delta \phi,\Delta \eta$) is the number of pairs at the corresponding
values of $\Delta \phi,\Delta \eta$ coming from the same event, after we have
summed over all the events. M($\Delta \phi,\Delta \eta$) is the number of pairs
at the corresponding values of $\Delta \phi,\Delta \eta$ coming from the mixed
events, after we have summed over all our created mixed events. A mixed event
pair has each of the two particles chosen from a different event. We make on
the order of ten times the number of mixed events as real events. We rescale
the number of pairs in the mixed events to be equal to the number of pairs in
the real events. 

The behavior of the HBT quantum interference radii\cite{HBT}, especially
$R_s$ was interpreted as indicative of observation of spatial radii 
$\sim$2fm in RHIC central Au + Au collisions at $\sqrt{s_{NN}}$ = 200 GeV
for $p_t$ $>$ 0.8 GeV/c. $R_s$ reduced to $\sim$2fm from $\sim$6fm as
$p_t$ increased from $\sim$ 0.2 GeV/c to greater than $\sim$ 0.8 GeV/c.
HBT quantum interference is by necessity only measured for 
pairs of charged particles of the same sign with small difference 
in momentum. In our model all particles coming from a bubble are constrained 
to come from a $\sim$ 2fm radius. Particles above 0.8 GeV/c $p_t$ that come 
from different bubbles differ in momentum such that they do not show 
HBT quantum interference.

The generally accepted explanation for observing these increasingly smaller 
final state HBT radii as $p_t$ increased was that radial flow increasingly
focused the viewed region of the overall final state source into a smaller
volume as $p_t$ increased. We refer to this phenomenon as phase space
focusing due to flow. We realized (in our first bubble model 
paper\cite{themodel} published in 2003) that flow phase space focusing
implies that the viewed region on the surface for $p_t$ $>$ 0.8 GeV/c of 
the fire ball would have a volume with radii of $\sim$2fm. With in this 
volume a hot source producing a larger number of particles would move out 
away from the surface, and being focused together would lead to an increase of
particles emitted in an angular region. However the HBT correlation function
has the property that a ring of essentially similar bubbles as assumed
in our PBM model would image on top of each other forming an average 
bubble which HBT would be viewing. The lower $p_t$ cut of 0.8 GeV/c would 
allow HBT to view and resolve this average bubble formed from differences in 
momentum of two particles for $p_t$ above 0.8 GeV/c. Our model 
populated these HBT viewed regions with a ring of bubbles or gluonic
hot spots producing a larger number of particles with an angular correlation.
We form a correlation function based on the difference of angles between two 
charged particles which images the 12 bubbles on top of each other. 

It is important to note that we have calculated that the correlations
observed at RHIC are strong enough so that if there was only one bubble
instead of a ring of bubbles distributed around the azimuth as our PBM 
assumed, one would produce an angular region with huge amplitude spikes
in individual events. These spikes are not observed in individual events
at RHIC. Therefore the RHIC correlation data has to be built of smaller
distributed correlated regions as assumed by the PBM.
 
We had used a virtually identical bubble ring and the same correlation 
function in our first bubble model\cite{themodel} which was published in 2003
and was able to predict and subsequently explain important general 
characteristics of the experimental analysis of the two charged particle 
correlation data for central collisions\cite{centralproduction}. This 
experimental paper used the $p_t$ cuts developed in our 2003 paper 0.8 $<$ 
$p_t$ $<$ 2.0 GeV/c. The 2 GeV/c cut was employed to make jet contamination 
negligible. The angular correlations used the differences of the azimuthal 
angles ($\Delta \phi$) and the differences of psuedorapidity ($\Delta \eta$) 
of all charge particle pairs which had the property that imaged all bubbles
in the ring on top of each other. This allowed the phase space focusing by 
radial flow to provide consistency with the observed correlation data 
analyses and the HBT observations.

An important assumption made in our PBM is that in a central heavy ion
collision (e.g. Au + Au) at high RHIC energies a high density of energetic 
partons (virtually all gluons) form a dense opaque fireball. This dense 
opaque fireball has a large amount of radial flow and can be described 
very well by a blast wave model. The usually employed blast wave model we 
used\cite{HBT} has its maximum velocity at the surface of the fireball at 
kinetic freezeout of approximately (3/4)c.

A theoretical pQCD calulation\cite{Kajantie} in 1987 concluded that jets 
formed with initial parton transverse momenta of around 3 GeV/c (also
applicable down to 2 GeV/c) would become thermalized in a $\sqrt{s_{NN}}$ = 
200 GeV U + U collision at RHIC and would not escape from the system. 
Therefore these jets would not result in the correlations observed at 
RHIC\cite{centralproduction}. One could speculate that these thermalized
jets form the dense opaque fireball.

There is direct experimental evidence of strong quenching of high $p_t$
particles at RHIC (e.g. Refs.\cite{quench1,quench2,quench3}). An experimental
result which is independent of our model demonstrates that in the central 
region charged particles with $p_t$ of (0.8 $<$ $p_t$ $<$ 4.0 GeV/c) are 
emitted from the surface of the dense opaque fireball. This point was 
demonstrated in Section IV B of Ref.\cite{centralitydependence}. We quote
the last 3 sentences from this Section IV B which states the conclusion.
``This surface or near surface hadronization and emission from the fireball 
both occur in the central region and all other centralities where there is
appreciable particle density. In the most peripheral bins the particle 
density is low enough to allow undisturbed fragmentation and thus no change 
in the CD correlation. Thus the CD behavior is consistent with a surface 
emission model such as Ref.\cite{PBM}.'' This result implies that surface
or near surface emission occurs for charged particles with  $p_t$ 
(0.8 $<$ $p_t$ $<$ 4.0 GeV/c) for all centralities.

The HIJING event generator\cite{hijing} combines Pythia jets\cite{pythia}
and the Lund model\cite{lund} and thus was a familiar base for constructing 
our PBM model. As we had done in our earlier bubble model\cite{themodel}, 
we replaced the Pythia jets in HIJING with our bubble ring. Momentum, 
energy and charge conservation are all satisfied within the bubble ring 
in the PBM. The bubble ring becomes the source of emmitted particle correlation
generated by Pythia which we used for fragmentation of the bubbles. We made 
the approximation that hard jet particles are essentially removed or have
their correlations removed by quenching. The only remaining particles from
HIJING are the beam jet fragmentation particles which are soft and have no 
correlation. These particles become our background particles in the PBM.
However we did include the effects of elliptic flow\cite{ellipticflow}
on the soft beam jet fragmentation particles, since elliptic flow does 
generate a small cos$(2\Delta \phi)$ term in the correlation even in the 
central collisions. The procedure we employed was to in each event determine 
the reaction plane and modulate the soft beam fragmentation particles by
the elliptic flow term $2 v_2$ cos$(2\Delta \phi)$ which was a sufficient
approximation for elliptic flow effects. Thus the elliptic flow effects were 
put into the model on an event by event basis. In the PBM we used angular 
correlations of charge particle pairs to predict and fit the experimentally 
determined correlations\cite{centralproduction}.

In Fig.3 of the PBM publication\cite{PBM} we schematically show a number
(3-4) of parallel parton showers contained in each bubble. The parton showers
must be parallel to give results which are consistent with the experimental
analysis of angular correlations.

The particle production from our bubbles uses a similar parton QCD shower
fragmentation as a jet with a well defined $\phi$ angle (Fig.2 of the 
PBM\cite{PBM}). The $p_t$ distribution of the partons inside the bubble is 
similar to pQCD but has a suppression in the high $p_t$ region like the
data\cite{quench1,quench2,quench3}. The 3-4 partons have different
longitudinal momentum. At kinetic freezeout we used Pythia fragmentation 
functions\cite{pythia} for the bubble fragmentation to form the final state
emitted charged particles.

Models that successfully predict and fit non-perturbative QCD experimental
results reasonably, almost always have to adjust some parameters when
comparing with experimental data. In the PBM there are two such adjustable
parameters which are the number of partons in a bubble and the longitudinal
momenta of the partons. Ever since Landau discovered in cosmic rays a long
time ago that excited nuclear fireballs exhibited a longitudinal
expansion this fact was well known. Adjusting the longitudinal momenta
of the partons is obviously necessary to explain the expansion in 
$\Delta \eta$ in the central production experiment\cite{centralproduction}.
The two parameters in the PBM were adjusted for a set number of bubbles by 
comparing the PBM fit to the experimentally determined final state charge pair
correlation (CI defined below).

In our models (e.g. PBM) the experimental correlation data analysis we are
comparing to utilizes the correlations of charge particle pairs. There are
2 types of such correlated pairs, namely unlike-sign charge pairs (US) and
like -sign charge pairs (LS). The total sample of correlated charge pairs 
generated and emitted in the final state of a theoretical model is equal to
the average of the US and LS correlations ((US + LS)/2). In an experiment
the total sample detected depends on the acceptance and efficiency of the 
detector. One makes cuts on the theoretical model to account for these
effects. However the correlation function used in equation 1 is conventionally 
used in experimental analyses and is drastically independent of acceptance 
effects as stated in Section 2.1\cite{PBM} allowing reasonable comparisons.

The Charge Independent (CI) correlation is conventionally defined as the 
unlike-sign charge pair correlation (US) + like-sign charge pair correlation 
(LS). The total correlation derived when using all particle pairs independent
of what charge signs are used to form the correlation is equal to CI/2. Thus 
CI/2 gives the average structure of the correlated emitting sources
independent of charge and represents the overall physical phenomenon.

The Charge Dependent (CD) correlation is  conventionally defined as the 
US - LS. The subtraction of the total like-sign charge pairs correlation (LS)
in forming the CD is equivalent to removal of the opposite sign charge pairs
which are not from the same space-time region where charge has to 
balance\cite{balfun,centralproduction}. Therefore the CD is expected to
represent the correlation of unlike-sign pairs from the same space-time
region where charge is balanced as modified by interaction with the medium.
In our surface emission model (PBM) the interaction with the medium is
absent therefore, the CD is expected to exhibit the correlation of Pythia
jets which are produced in a vacuum.
  
In Section 4.2 of the PBM\cite{PBM} we compare the total CI correlations of
PBM and precision experimental correlation data\cite{centralproduction}.
We adjusted the two parameters by comparing the CI of the PBM with the CI of 
the data. We show the PBM and the data agree within less than 10\% of the
total CI correlation which is a reasonably quantitative agreement.

In Section 4.3 of the PBM\cite{PBM} we compare the CD correlations of
PBM with the experiment by making use of the fact that the net charge 
fluctuation suppression is directly related to an integral over the CD.
Thus we compared the net charge fluctuation suppressions of the data with the 
PBM and found agreement within errors. We did the comparison of the CD in 
this way since the experimental paper we compared the model with considered 
this to be an important aspect of the CD correlation and chose to treat the 
CD in this manner.

\section{Extension of PBM to cover 0-80\% centralities}

The PBM is a successful bubble model for central heavy ion (e.g. Au + Au)
collisions at the highest energy at RHIC as discussed in the previous 
section. However an interesting question that arises is what would
happen to bubble production and the general characteristics of charge pair
correlations as centrality varies from most central (impact parameter near 
zero) to peripheral collisions. The PBM has been successfully tested in the 
centrality range approximately (0-10\%). A new precision RHIC minimum bias 
trigger data analysis for  Au + Au collisions at $\sqrt{s_{NN}}$ = 200 GeV in 
transverse momentum range 0.8 GeV/c to 4.0 GeV/c covers the 0-80\% centrality
range\cite{centralitydependence} and is ideally suited for investigating the
varying centrality evolution of the PBM. This experimental analysis was done
in a manner which was a logical extension of the central production
paper\cite{centralproduction}. The data, the charge particle pairs
correlations (US and LS), the CI and the CD were treated in a similar manner, 
but analyzed and fit separately in each of nine centrality bins.

In the data analysis\cite{centralitydependence} the most central bins US, LS, 
and CI were consistent with the results of the prior central production 
experiment. However as one moves from central to peripheral bins a jet-like 
component is increasingly evident (in the data analysis) till the most 
peripheral bins where there is only jet-like correlations. The elliptic flow 
amplitude $2 v_2^2$ as part of the correlation increases as one moves to 
more peripheral centralities. This flow reaches its maximum at 40-50\% 
centrality (see Fig. 28 of Ref.\cite{centralitydependence}).

We will proceed to extend the PBM to include the entire 0-80\% centrality 
region. We name this extended version of the PBM as PBME. Several obvious
characteristics of the experimental and theoretical analysis that must be 
included in this extension of the model to PBME are discussed below.
 
The previous central collision bubble ring geometry was well suited for the 
most central collisions situation since the highest energy densities are
circular around the beam which is the expected geometric symmetry. As one
moves from central toward peripheral the decreasing size and change of 
shape of the overlap region of the two Au nuclei determines where the
energy densities are highest. The overlap of matter in the two Au nuclei
becomes greater in the reaction plane region, while the overlap of matter 
becomes less outside of the reaction plane. This breaking of symmetry 
will modify the overall spatial shape and location of the bubbles.

The effects of elliptic flow\cite{ellipticflow} on the events in each 
centrality bin were put into the model using the same procedure we used as 
previously described in Section II for the PBM. The procedure we employed was 
to in each event determine the reaction plane and modulate the soft beam 
fragmentation particles by the elliptic flow term $2 v_2$ cos$(2\Delta \phi)$ 
which was a sufficient approximation for elliptic flow effects. Thus the 
elliptic flow effects were put into the model on an event by event basis. 

Jet quenching is largest in the most central collisions and decreases as one 
moves to more peripheral bins. We found a sufficient way to put the effects 
of strong jet quenching in the central collisions. We set jet quenching to
its maximum in HIJING for 0-30\% centralities by removing all jets 
(quenched away). For centralities 30-80\% we use the non-jet-quenching 
version of HIJING thus all jets become part of the event. The soft beam jet
particles have elliptic flow as described above.

We relied on the blast wave to determine the geometry of where the energy
density is highest. These regions of high energy density is where bubbles
are formed. In the final state at kinetic freezeout the bubbles are located 
on the fireball (blast wave) surface which is the source of emmitted correlated
charge particle pairs generated by Pythia fragmentation of the bubbles.
Thus the blast wave determined the location, number and geometry of the
bubbles.

\section{General characteristics of the PBME}

For the most central bin we have the same bubble geometry and partons per 
bubble as Ref.\cite{PBM}; namely 12 bubbles and 3-4 partons per bubble. 
As we move away from the most central the number of partons per bubble 
decreases dropping to 3 then 2 and finally 1 at 50-60\% centrality. 
The number of bubbles formed per event decreases from 12 in the most central 
bin to 0.3 in the 50-60\% bin (see Table I in this section). 
 
The most central collisions have a bubble ring symmetry about the beam axis 
(see Fig.1 of Ref.\cite{PBM}). The region of highest energy density 
will also be symmetric about this axis and the ring of bubbles is the expected 
geometry. As we move to more peripheral collisions the symmetry becomes 
defined by the reaction plane. The region of highest energy density
becomes more concentrated in the region of the reaction plane. Therefore it is 
reasonable to expect the ring symmetry will be broken and the bubbles farthest 
from the reaction plane would disappear. We find comparing 0-5\% centrality to 
5-10\% centrality the 12 bubbles reduce to 10 bubbles with the bubbles 
perpendicular to the reaction plane gone. For 10-20\% the number drops to 7 
bubbles with 4 bubbles being near the reaction plane and 3 appearing above and 
below the plane, while for 20-30\% we drop to 5 bubbles with 4 bubbles being 
near the reaction plane with only 1 appearing above or below the plane. 
Moving to 30-40\%, 40-50\%, and 50-60\% the bubbles are in the reaction plane 
region with the probability per event of making a bubble being 1.5, 0.6, 0.3 
respectively (see Table I). The above changes in bubble production are due to 
the decrease in energy density as the \% centrality range becomes more
peripheral which decreases the overlap region of the Au + Au colliding nuclei. 

The STAR experiment has measured charged particle pair correlations for 
minimum bias Au + Au events at $\sqrt{s_{NN}}$ = 
200 GeV\cite{centralitydependence}. The $p_t$ range of that data is 0.8 
to 4.0 GeV/c for the entire $0-360^\circ$ $\phi$ range and the $\eta$ range 
$|\eta| < 1.0$. Both the experimental data analysis compared to and the PMBE 
model utilize two dimensional $\Delta \phi$ $\Delta \eta$ correlations. 
See examples of these two dimensional perspective plots in 
Refs.\cite{PBM,centralproduction} and Fig.11-12 (this paper).

In order to compare and present these two dimensional plots for the new 
data\cite{centralitydependence} and the PBME we divided the entire 
$\Delta \eta$ region into five $\Delta \eta$ bins which covered the entire 
$\Delta \eta$ range. Each $\Delta \eta$ bin could then be presented as a one 
dimensional projection and a comparison between data and model can be made. 
The five $\Delta \eta$ bins were 0.0 to 0.3, 0.3 to 0.6, 0.6 to 0.9, 0.9 to 
1.2 and 1.2 to 1.5. In each of these five $\Delta \eta$ bins the $\Delta \phi$ 
correlations for the charged particles covered the entire $\Delta \phi$ range 
0 - $180^\circ$. Due to demonstrated symmetry in the data and the model the 
$360^\circ$ $\Delta \phi$ range which was experimentally detected was folded 
resulting in the $180^\circ$ ranges. 

\begin{center}
\begin{tabular}{|c|r|r|}\hline
\multicolumn{3}{|c|}{Table I}\\ \hline
Centrality & bubbles per event & partons per bubble \\ \hline
$0-5\%$ & 12 & 3-4 \\ \hline
$5-10\%$ & 10 & 3  \\ \hline
$10-20\%$ & 7 & 3  \\ \hline
$20-30\%$ & 5 & 3  \\ \hline
$30-40\%$ & 1.5 & 2  \\ \hline
$40-50\%$ & 0.6 & 2  \\ \hline
$50-60\%$ & 0.3 & 1  \\ \hline
\end{tabular}
\end{center}                                     

\bf Table I. \rm Parameters of bubble model with centrality.

\section{Charge Independent (CI) correlation}

In this section we are going to compare the PBME model predicted CI
correlations with the STAR experimental analysis 
results\cite{centralitydependence}. The two particle correlations are 
formed from two different types of charge particle pairs:

1) Unlike-Sign charge pairs (US)

2) Like-Sign charge pairs (LS)

The Charge Independent (CI) total correlation = US + LS. It is the sum of the
CI signal + Background correlations in the final state after kinetic freezeout.
Thus it is the total two particle correlation observed in the detector 
(STAR TPC). The entire CI correlation (signal + background) is used for 
comparing the analysis results with the model. This eliminates any model
dependence on the separation of signal from background. Certain necessary 
corrections and cuts in the experimental analysis were applied to the PBME 
model (this paper) so that a quantitative comparison could be made with 
the experimental analysis. The CI displays the average correlation 
structure of the emitting sources in the final state after kinetic 
freezeout and thus is physically significant.

The five plots comparing the experimental analysis of the CI and the PBME fits 
are shown in Fig. 1-5. On the vertical axis the CI is multiplied by the 
average event multiplicity within the particular centrality bin shown by the 
symbols on the plot. This procedure is necessary when one compares different 
centralities in order to make the comparison independent of multiplicity.

The correlation function is given in equation 1. The experimentally 
observed correlation has a numerator which is proportional 
to the number of correlated particle pairs which itself is 
proportional to the multiplicity. However the denominator is 
proportional to the total number of pairs which can be formed 
which is proportional to the square of the multiplicity. 
Therefore in order to make comparisons of different centralities 
or other experiments one multiplies by the average multiplicity to remove the
dependence on the multiplicity. This procedure is referred to as multiplicity 
scaling or multiplicity scaled.  

For each centrality we generated 500,000 simulated events with the bubble 
geometries presented above. The impact parameter range of HIJING for the 
different centralities is given in Table II. See Table I for the number of 
bubbles per event and partons per bubble as a function of \% centrality bins.

\begin{center}
\begin{tabular}{|c|r|}\hline
\multicolumn{2}{|c|}{Table II}\\ \hline
Centrality & Impact parameter (fm) \\ \hline
$0-5\%$ & 0.0-2.9 \\ \hline
$5-10\%$ & 2.9-4.1 \\ \hline
$10-20\%$ & 4.1-5.8  \\ \hline
$20-30\%$ & 5.8-7.1  \\ \hline
$30-40\%$ & 7.1-8.2  \\ \hline
$40-50\%$ & 8.2-9.2  \\ \hline
$50-60\%$ & 9.2-10.1  \\ \hline
$60-70\%$ & 10.1-10.9  \\ \hline
$70-80\%$ & 10.9-11.7  \\ \hline
\end{tabular}
\end{center}                                     

\bf Table II. \rm Impact Parameters of HIJING with Centrality.

In Fig.1 to Fig.5 we used the STAR calculated Charge Independent (CI) 
correlation which is the average correlation of the unlike-sign charge pairs 
correlation plus the like-sign charge pairs correlation. In order to allow
comparison with the different centrality bins we and also STAR multiplied 
by the multiplicity, because as explained previously this removes the dilution 
of the signals due to the quadratic increase of pair combinations. The 
multiplicity that is used in Ref.\cite{centralitydependence} is based 
on the particles measured in the STAR TPC (Time Projection Chamber). There 
are readout boundaries between the 12 TPC sectors which cover the azimuth that 
do not measure tracks. We put these readout boundaries in our Monte Carlo 
generation of our particles, which cause a loss of approximately 10\% of the 
particles. 

The multiplicity scaled correlation for each centrality for a given 
$\Delta \eta$ bin is plotted with the maximum angle of the away side 
($\Delta \phi = 180^\circ$) shifted to the same value. The horizontal line 
for each centrality shows the shifted average multiplicity line which was 
normalized to a mean of 1 in the original CI correlation and became 
equal to the average multiplicity after becoming rescaled to the scaled 
correlation. The solid curves are the PBME calculations shifted by the 
same amount as the data. The agreement between the PBME and the RHIC data in 
Fig.1 to Fig.5 is within a few percent of the total correlation. Considering 
that the model (PBME) does not completely include important non-perturbative 
QCD effects contained in the data we consider this a reasonable quantitative 
agreement. The CI x multiplicity displays the average structure of the 
correlation sources at kinetic freezeout.

%\vspace{-0.5cm}
\begin{figure*}[ht] \centerline{\includegraphics[width=0.800\textwidth]
{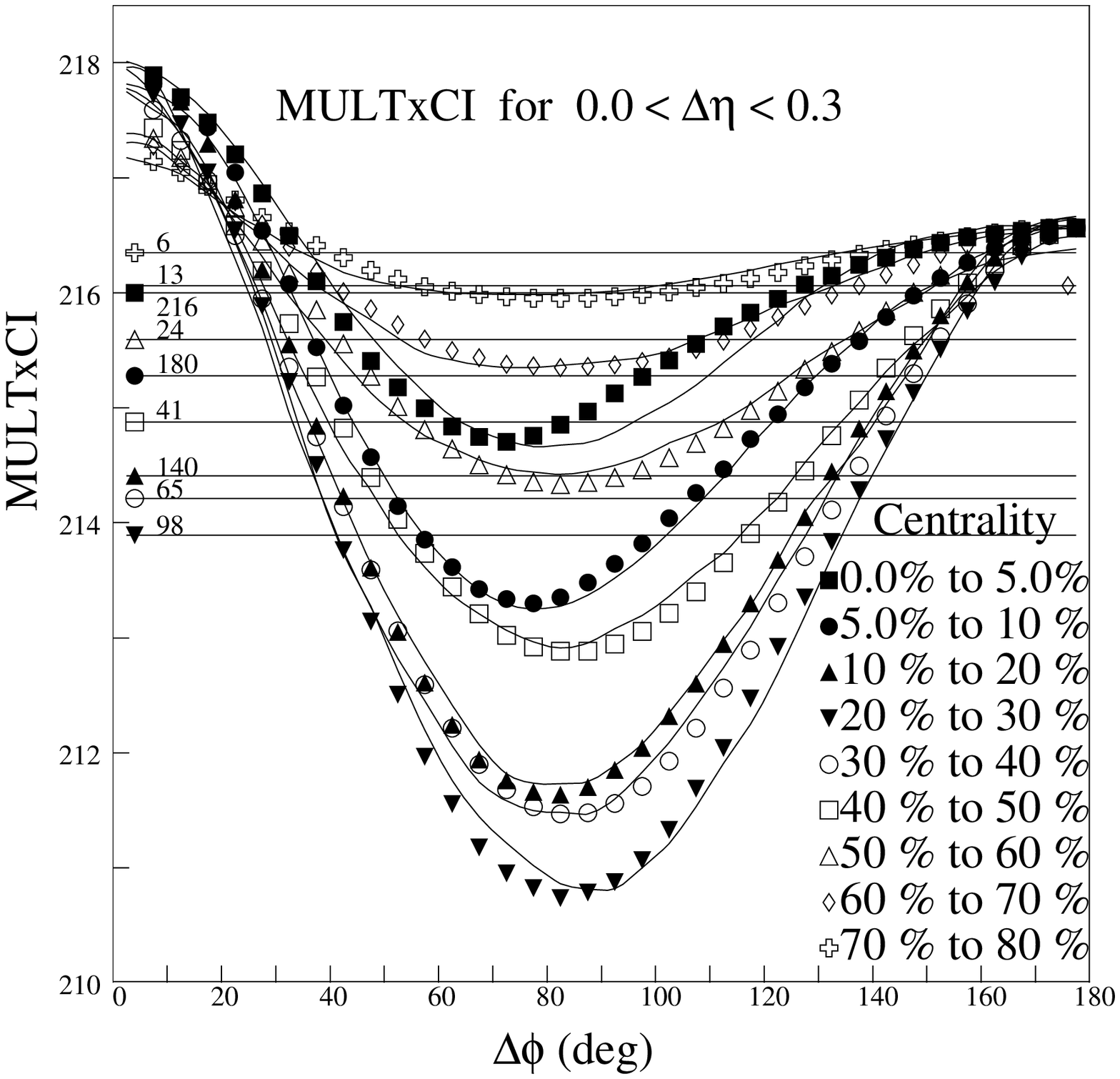}} \caption[]{The multiplicity(MULT) times the CI correlation vs. 
$\Delta \phi$ for 0.0 $< \Delta \eta <$0.3. Nine centralities are shown 
from 70\% to 80\% with centrality increasing to 0\% to 5\%. 216 is the 
multiplicity for 0\% to 5\% and all other centralities are shifted 
up so that the $180^\circ$ value is equal. Each multiplicity for 
each centrality is shown shifted. The solid curves are the 
bubble model (PMBE) calculations shifted by same amount as the 
data. The PBME and RHIC data agree within a few percent of the 
total CI correlation in each centrality bin in each $\Delta \eta$ 
range for all $\Delta \phi$ angles.}
\label{figure1}
\end{figure*}
                                                                               
%\vspace{-0.5cm}
\begin{figure*}[ht] \centerline{\includegraphics[width=0.800\textwidth]
{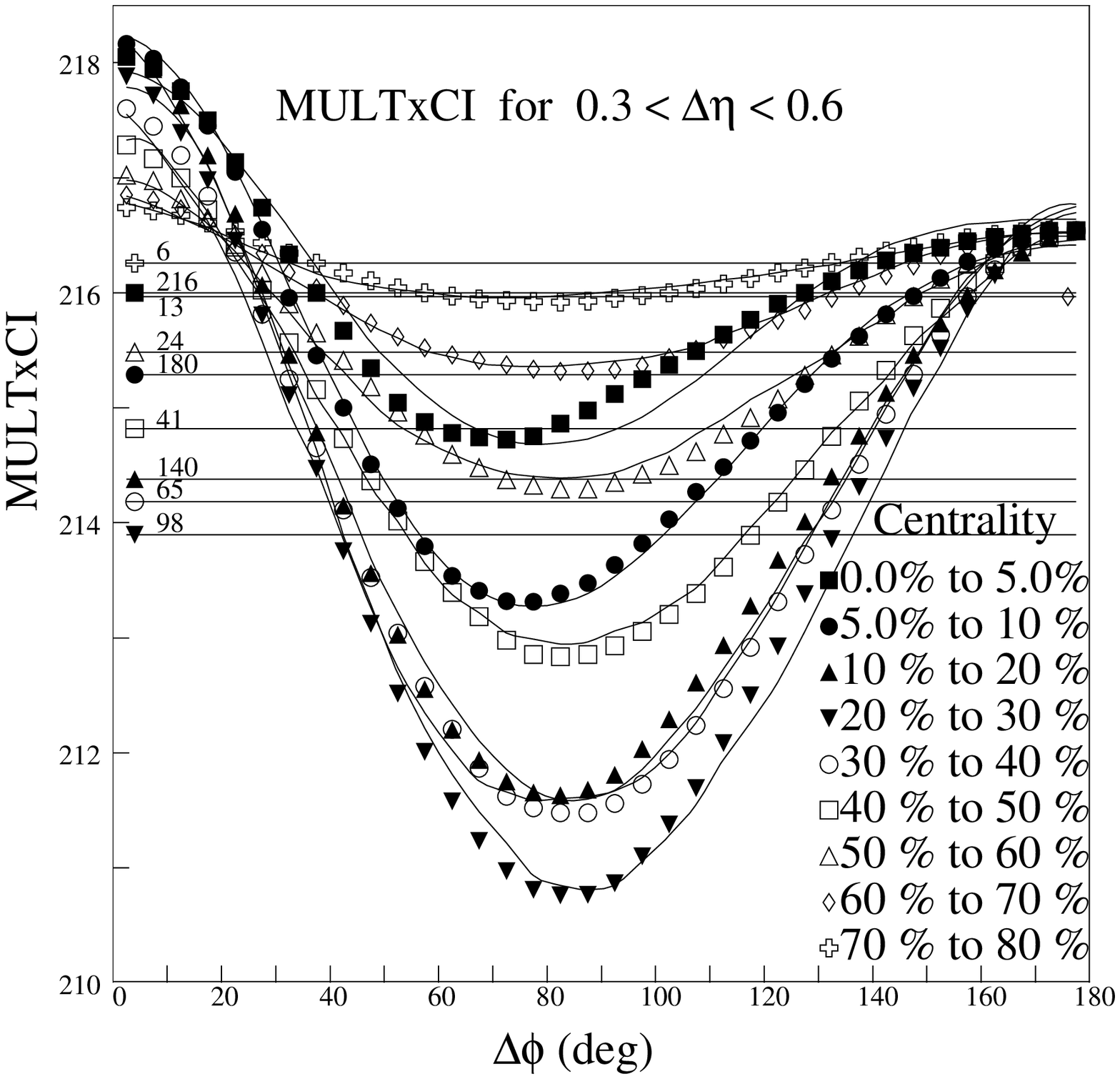}} \caption[]{The multiplicity(MULT) times the CI correlation vs. 
$\Delta \phi$ for 0.3 $< \Delta \eta <$0.6. Nine centralities are shown from 
70\% to 80\% with centrality increasing to 0\% to 5\%. 216 is 
the multiplicity for 0\% to 5\% and all other centralities are 
shifted up so that the $180^\circ$ value is equal. Each multiplicity 
for each centrality is shown shifted. The solid curves are 
the bubble model (PBME) calculations shifted by same amount as 
the data. The PBME and the RHIC data agree within a few percent for 
the total CI correlation}
\label{figure2}
\end{figure*}
                                                                             
%\vspace{-0.5cm}
\begin{figure*}[ht] \centerline{\includegraphics[width=0.800\textwidth]
{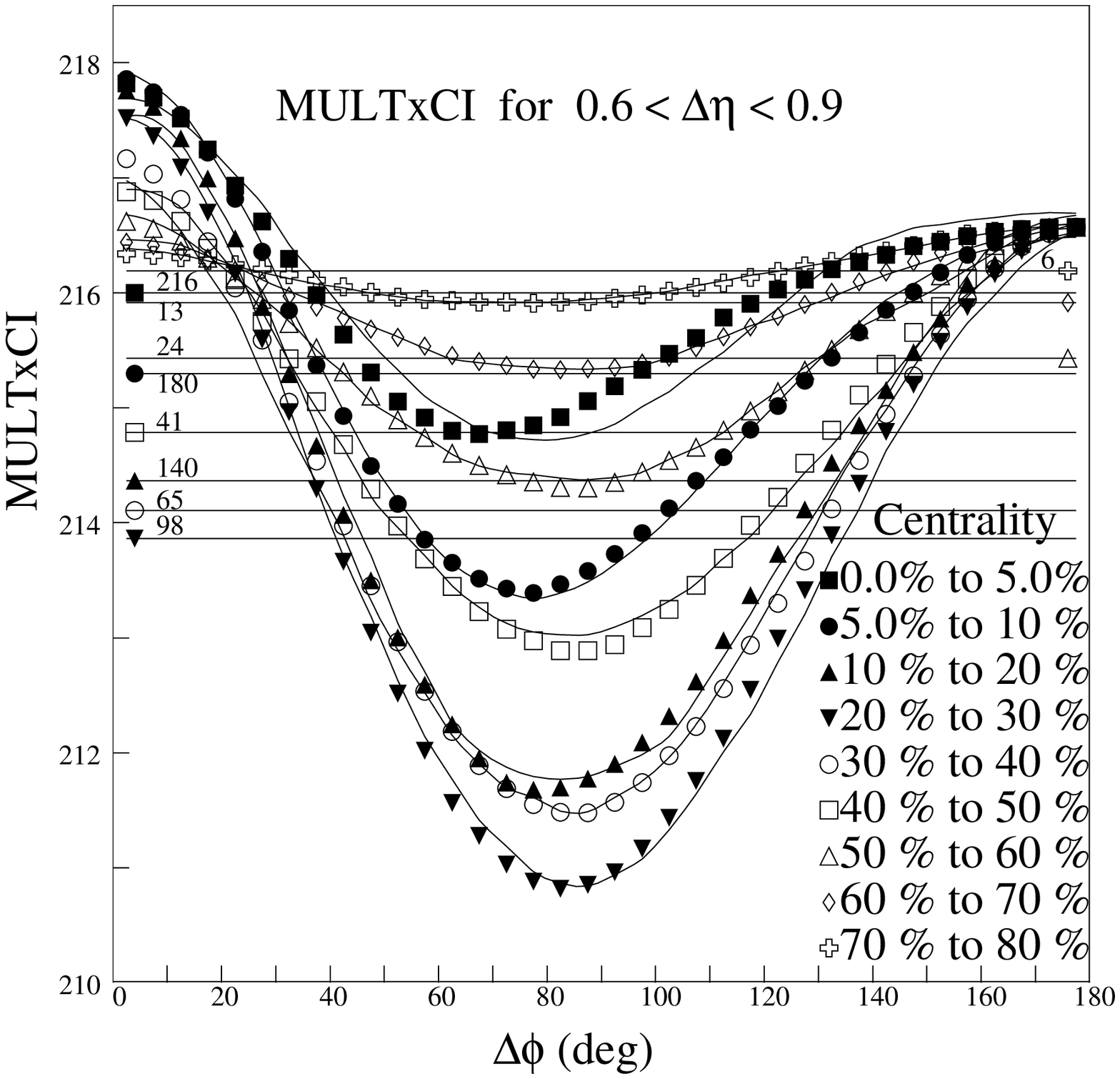}} \caption[]{The multiplicity(MULT) times the CI correlation vs. 
$\Delta \phi$ for 0.6 $< \Delta \eta <$0.9. Nine centralities are shown 
from 70\% to 80\% with centrality increasing to 0\% to 5\%. 216 is 
the multiplicity for 0\% to 5\% and all other centralities are 
shifted up so that the $180^\circ$ value is equal. Each multiplicity 
for each centrality is shown shifted. The solid curves are 
the bubble model (PBME) calculations shifted by same amount as 
the data. The PBME and the RHIC data agree within a few percent for 
the total CI correlation}
\label{figure3}
\end{figure*}
                                                                             
%\vspace{-0.5cm}
\begin{figure*}[ht] \centerline{\includegraphics[width=0.800\textwidth]
{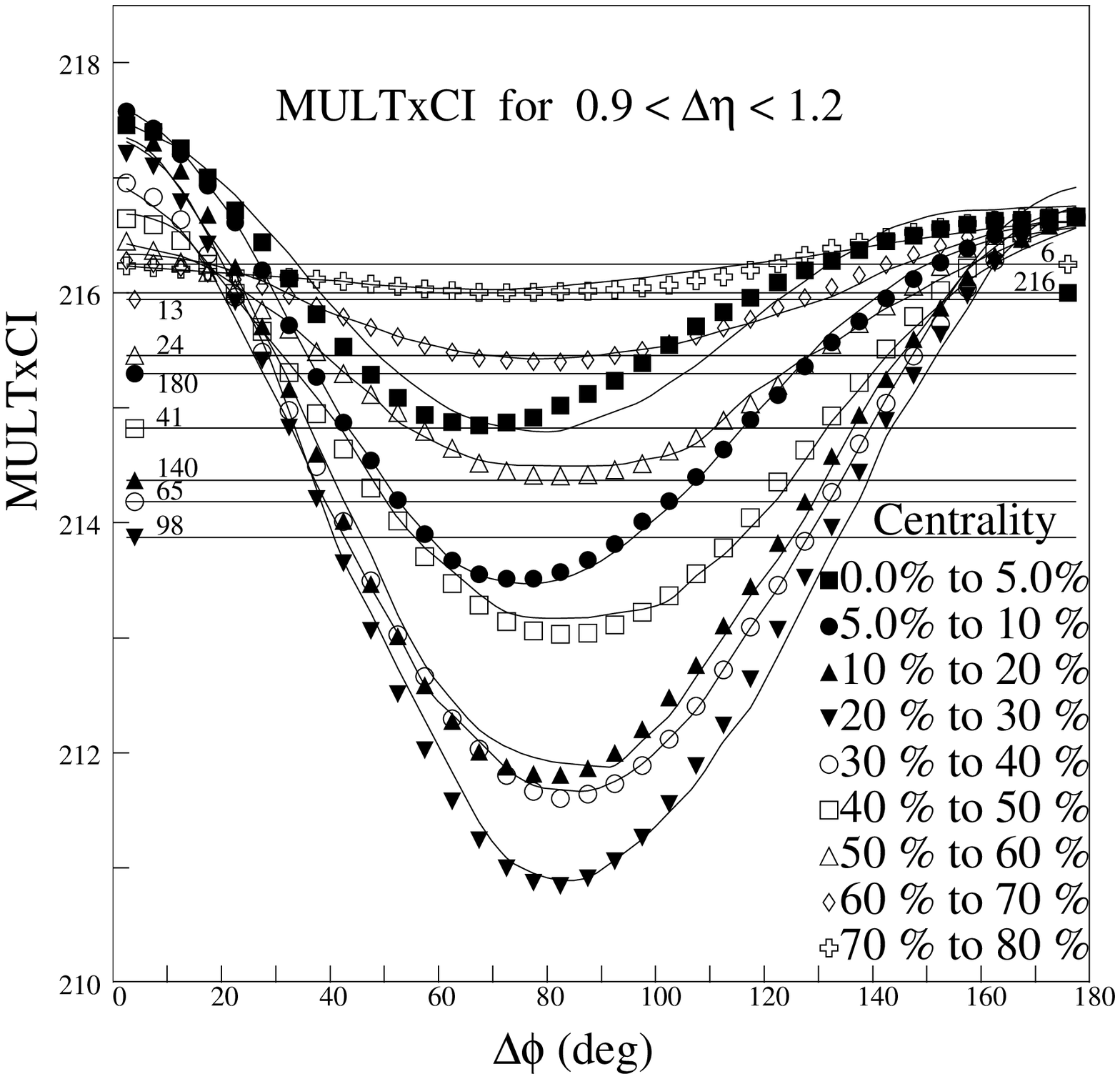}} \caption[]{The multiplicity(MULT) times the CI correlation vs. 
$\Delta \phi$ for 0.9 $< \Delta \eta <$1.2. Nine centralities are shown 
from 70\% to 80\% with centrality increasing to 0\% to 5\%. 216 is 
the multiplicity for 0\% to 5\% and all other centralities are 
shifted up so that the $180^\circ$ value is equal. Each multiplicity 
for each centrality is shown shifted. The solid curves are 
the bubble model (PBME) calculations shifted by same amount as 
the data. The PBME and the RHIC data agree within a few percent for 
the total CI correlation}
\label{figure4}
\end{figure*}
        
%\vspace{-0.5cm}
\begin{figure*}[ht] \centerline{\includegraphics[width=0.800\textwidth]
{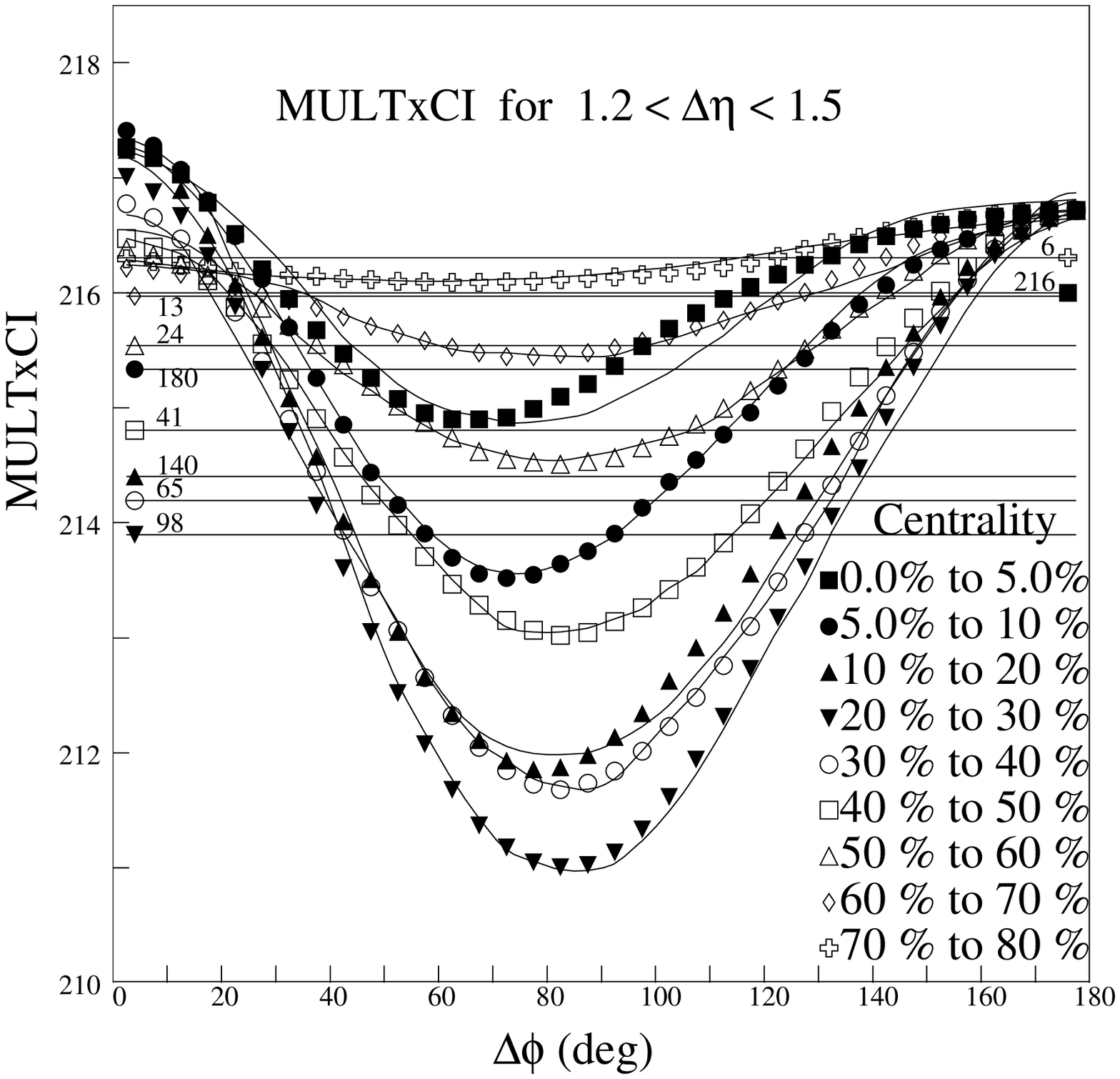}} \caption[]{The multiplicity(MULT) times the CI correlation vs. 
$\Delta \phi$ for 1.2 $< \Delta \eta <$1.5. Nine centralities are shown 
from 70\% to 80\% with centrality increasing to 0\% to 5\%. 216 is 
the multiplicity for 0\% to 5\% and all other centralities are 
shifted up so that the $180^\circ$ value is equal. Each multiplicity 
for each centrality is shown shifted. The solid curves are 
the bubble model (PBME) calculations shifted by same amount as 
the data. The PBME and the RHIC data agree within a few percent for 
the total CI correlation}
\label{figure5}
\end{figure*}
        
\section{Charge Dependent (CD) correlation}

We compare total experimentally observed CD correlations to total theoretically
predicted CD correlations in order to avoid possible uncertainties due to 
separation of signals and background.

The Charge Dependent (CD) correlation which is the difference between the 
unlike-sign charge pair correlations and the like-sign charge pair correlations
( US - LS ) displays a measure of the emission correlation of the opposite sign
pairs of particles emitted from the same space-time region at the time of 
hadronization\cite{balfun,centralproduction}. The CD as all correlations
in all centrality ranges in the PBME (which includes a HIJING jet 
component) has it's fragmentation of all partons determined by Pythia 
fragmentation\cite{pythia}. We use the same projection method of the 2-D
CD correlation into the same five $\Delta \eta$ ranges and multiply the CD
by the multiplicity as described and discussed in the previous Section V.
In Fig.6 we show the multiplicity times the CD $\Delta \phi$ correlation 
from Ref.\cite{centralitydependence} within  the $\Delta \eta$ range 
0.0 to 0.3 for each centrality compared to the bubble model (PBME) 
calculations. We achieve a good agreement between data and model. Similar 
results and good agreement with the PBME occurs for the four other 
$\Delta \eta$ ranges not shown. Thus we have good agreement in all 5 
$\Delta \eta$ ranges which together comprise the entire CD.  

In Fig.7 we plot from Ref.\cite{centralitydependence} the CD 
$\Delta \phi$ correlation for 4 $\Delta \eta$ bins covering the range 
0.0 $< \Delta \eta <$ 1.2.\footnote{The $\Delta \eta$ range 1.2 to 1.5 
(not shown) is just a flat background with very little CD signal from 
Pythia parton fragmentation left.} Each of 9 centralities shown were scaled 
so that the 5 - $10^\circ$ $\Delta \phi$ bin for 0.0 $< \Delta \eta <$ 0.3 
range is normalized to 1. This was done in order to remove the scale 
difference between the different centrality ranges and allow us to show 
that the CD shape is approximately independent of centrality. 
In Fig.7 the experimental analysis points in each $\Delta \eta$ 
range cluster around the four lines which corresponds to each of the four 
$\Delta \eta$ ranges generated by Pythia jets\cite{pythia}. Pythia jets were 
used as the jets in HIJING and Pythia fragmentation was used in the parton 
bubble model\cite{PBM}. Thus the CD shape is in good agreement with
Pythia and is independent of centrality. Pythia jet CD correlations are the
initial correlations of opposite sign charge pairs from the same space-time 
region at the time of hadronization\cite{balfun,centralproduction}. The
fact that there is essentially no change in these correlations at the time of 
kinetic freezeout demonstrates that there is little or no further interaction 
of these opposite sign charge pairs with the fireball medium from the time of 
hadronization till the time of kinetic freezeout. Thus both hadronization and 
kinetic freezeout occur at or very near the surface of the expanding fireball 
at all centralities. Hence in general the fireball is dense and opaque at 
kinetic freezeout. One should note that in the peripheral bins there is very
little matter in the short path length from any point to the surface. This 
is consistent with surface emission from the fireball at kinetic freezeout or 
undisturbed fragmentation in those peripheral bins where the path length to 
the surface is small.

Both the original parton bubble model (PBM)\cite{PBM} and the present extension
of the model (PBME) construct surface bubbles which are boosted by the 
expanding fireball. These bubbles at freezeout give results that are consistent
with experimental correlation data\cite{centralitydependence}. Furthermore 
the PBME fits the observed CD correlation to within a few percent of 
the correlation at all centralities. The PBME produces the CD shape that is 
consistent with Pythia jet CD correlations. As previously shown in Section V
the PBME fits the observed CI correlations to within a few percent of the 
correlation. The US and LS correlations are linear combinations of the CI 
and CD. Therefore they are also fit to within a few percent of the
correlation. Thus all two charge particle pair correlations are reasonably 
quantitatively fit by the PBME (Parton Bubble Model Extended). One should 
note that the fireball in both the PBM and PBME is treated in a blast wave 
model with the bubbles forming on the surface and emitting their final 
state particles at kinetic freezeout. The fireball surface is moving at 
the maximum velocity at kinetic freezeout. 

%\vspace{-0.5cm}
\begin{figure*}[ht] \centerline{\includegraphics[width=0.800\textwidth]
{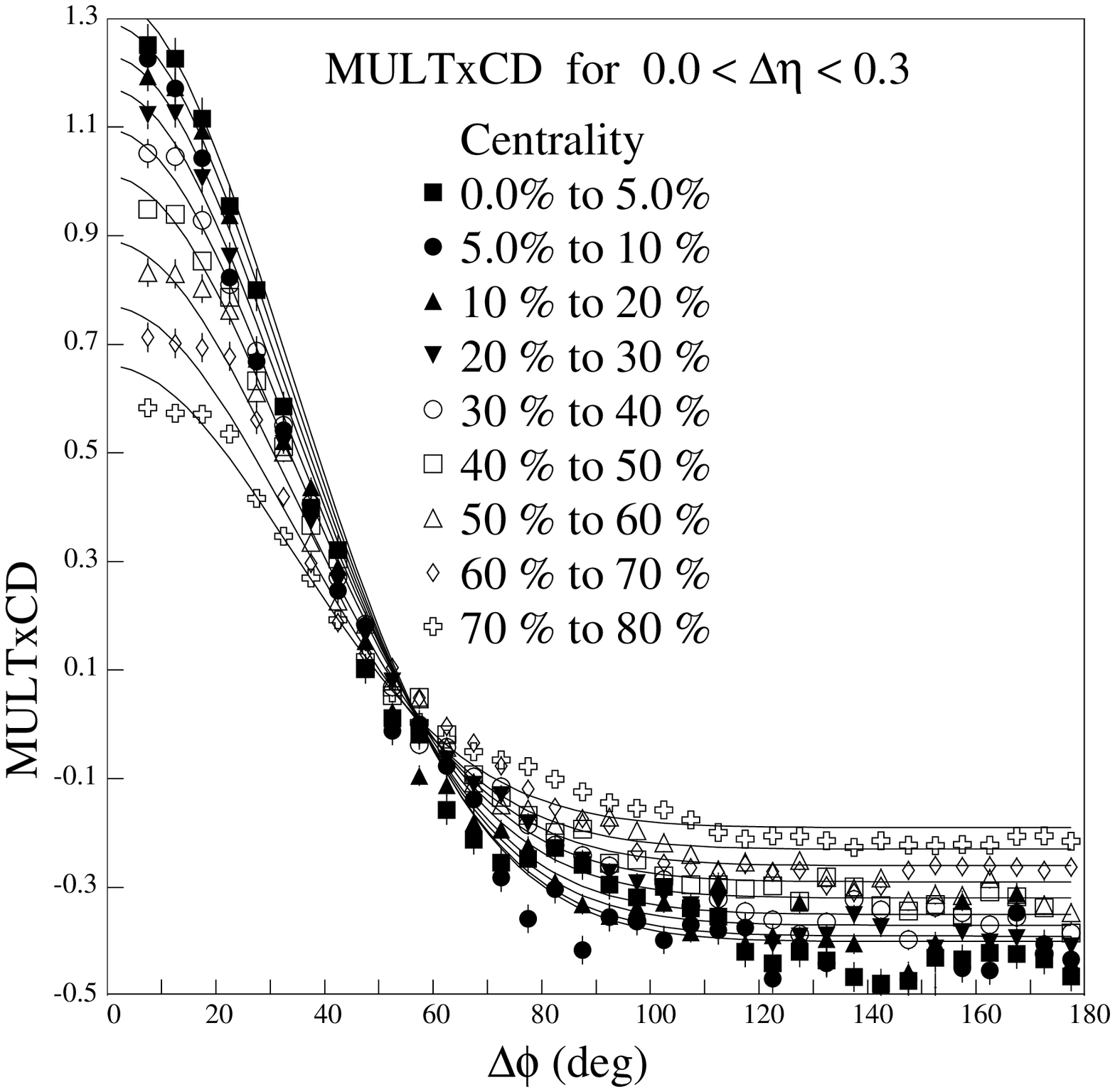}} \caption[]{The multiplicity(MULT) times the CD correlation vs. 
$\Delta \phi$ for 0.0 $< \Delta \eta <$0.3. Nine centralities are shown from 
70\% to 80\% with increasing centrality to 0\% to 5\%. The solid curves are 
the bubble model (PBME) calculations scaled by the multiplicity. We get 
good agreement.}
\label{figure6}
\end{figure*}
 
%\vspace{-0.5cm}
\begin{figure*}[ht] \centerline{\includegraphics[width=0.800\textwidth]
{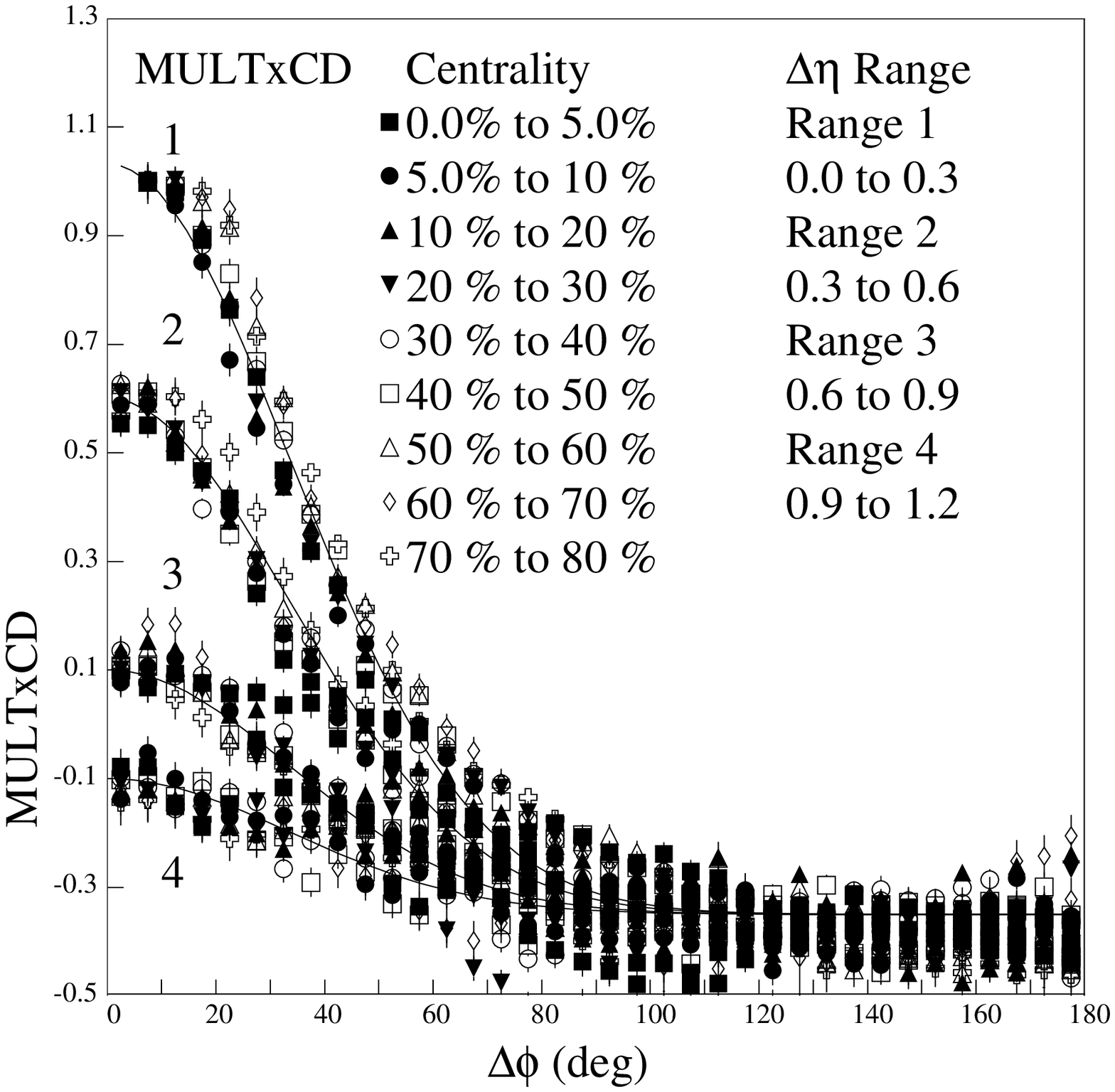}} \caption[]{The product of the multiplicity(MULT) and the 
CD correlation vs. $\Delta \phi$ for 4 $\Delta \eta$ bins. Nine centralities 
are shown from 70\% to 80\% with increasing centrality to 0\% to 5\%. Each of 
the centralities were scaled so that the 5 - $10^\circ$  $\Delta \phi$ bin for 
the $\Delta \eta$ range 0.0 to 0.3 is normalized to 1. The data points for 
each $\Delta \eta$ range cluster around the Pythia jet predictions (lines). 
Thus we see that the CD shape is approximately independent of centrality and 
the CD is approximately consistent with Pythia jets\cite{pythia} (line) at 
each centrality. The PBME fits the observed CD correlation to within a few 
percent of the correlation at all centralities since its predicted shape
is given by Pythia jets and we get a good agreement for multiplicity scaled
CD correlation for each of the centralities.}
\label{figure7}
\end{figure*}
 
\section{Further Discussion of the Parton Bubble Model Correlation}
 
We note from Tables I-III that as we move from the most central region (impact
parameter small) toward the peripheral bins the number of bubbles in a 
centrality bin per event and the number of partons per bubble both tend to 
decrease. In the two most peripheral bins bubble production is negligible.

\begin{center}
\begin{tabular}{|c|r|r|r|}\hline
\multicolumn{4}{|c|}{Table III}\\ \hline
No partons & particles per bubble & $p_t$ per bubble(GeV/c) & energy per bubble(GeV) \\ \hline
3-4 & 8.5 & 7.0 & 9.8 \\ \hline
3 & 7.0 & 5.7 & 8.0  \\ \hline
2 & 4.9 & 4.0 & 4.9  \\ \hline
1 & 2.6 & 2.1 & 2.1  \\ \hline
\end{tabular}
\end{center}                                     

\bf Table III. \rm Parameters of bubble model per number of partons.

\clearpage

Fig.8 and Fig.9 are 2 dimensional perspective plots -$\Delta \phi \Delta \eta$-
of the part of the CI correlation multiplied by the multiplicity 
(multiplicity scaled) which results from particles emmitted from the same 
bubble in the centrality bins shown. The $\Delta \phi$ angular distributions 
of the correlation are primarily confined to the near side ($\Delta \phi$ $<$
$90^\circ$). The shape of the $\Delta \phi$ correlation for 0-30\% centrality
bins peak near small angles and decrease as $\Delta \phi$ increases toward
$90^\circ$ and are qualitatively similar. However the $\Delta \eta$ 
distribution shape remains qualitatively broad in the 0-30\% centrality bins
exhibiting the $\Delta \eta$ elongation observed in the 
data\cite{centralitydependence}. One should note that there is very little 
away side ($\Delta \phi$ $>$ $90^\circ$) in any of these centrality bins. 

%\vspace{-0.5cm}
\begin{figure*}[ht] \centerline{\includegraphics[width=0.800\textwidth]
{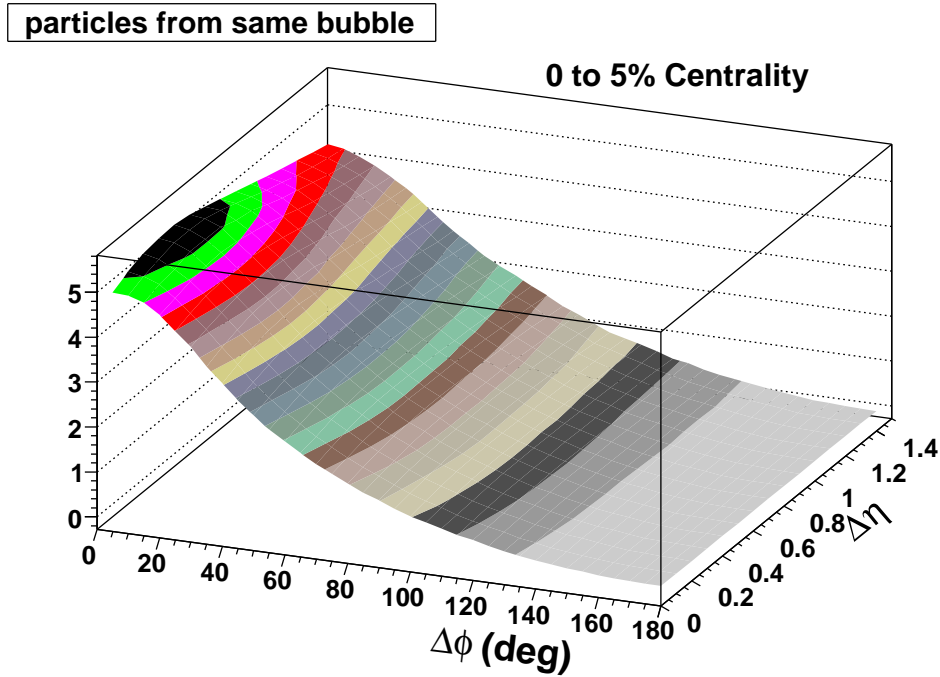}} \caption[]{``(Color online)'' The multiplicity(MULT) times 
the CI correlation for the 0-5\% centrality bin as a two dimensional 
$\Delta \phi$ vs. $\Delta \eta$ perspective plot. This is the part of the 
CI correlation that resulted from particles emitted from the same bubble.}
\label{figure8}
\end{figure*}

%\vspace{-0.5cm}
\begin{figure*}[ht] \centerline{\includegraphics[width=0.800\textwidth]
{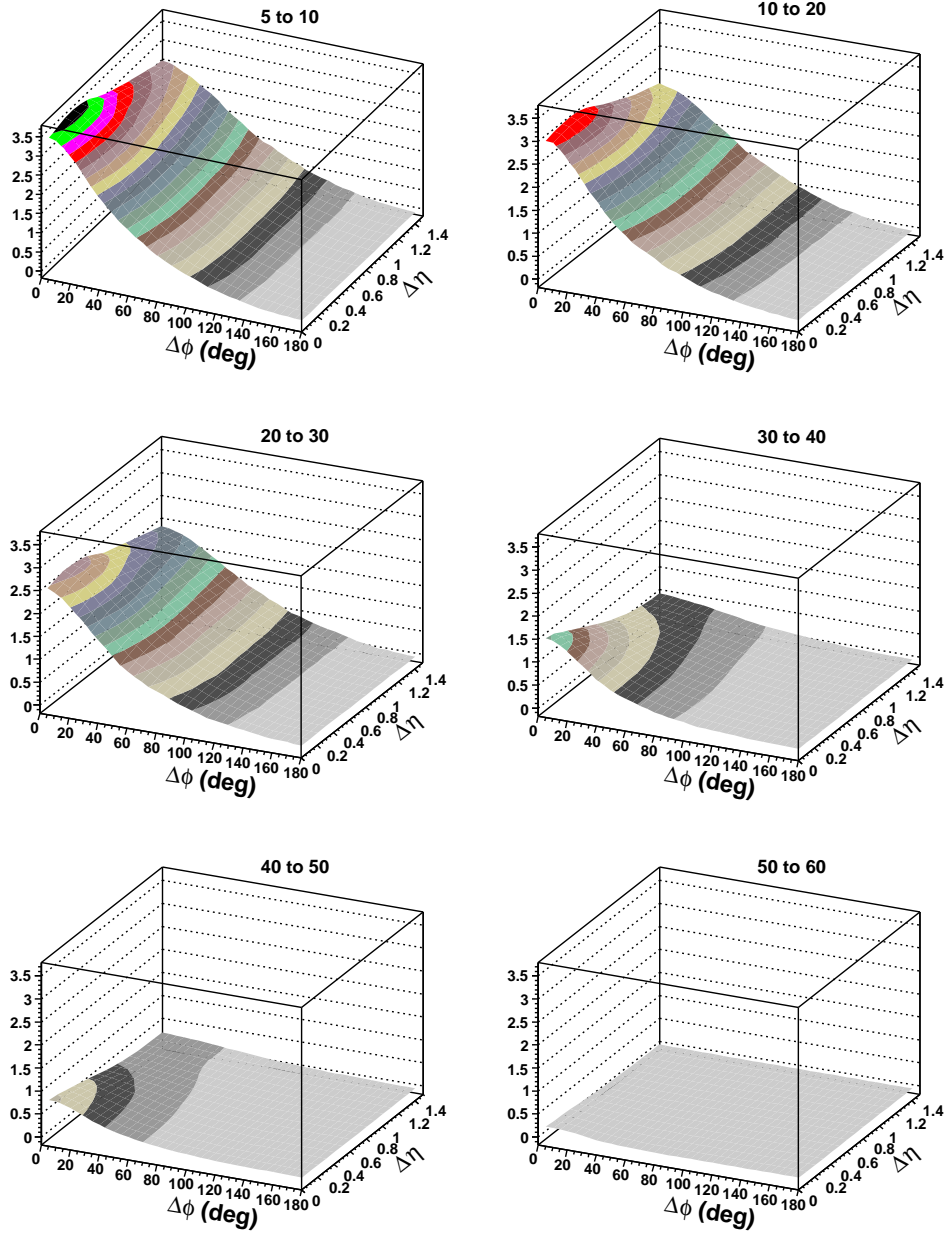}} \caption[]{``(Color online)'' The multiplicity(MULT) times 
the CI correlation for the six centralities as six two dimensional 
$\Delta \phi$ vs. $\Delta \eta$ perspective plots. This is the part of the 
CI correlation that resulted from particles emitted from the same bubble. 
We see in Fig.8 and Fig.9 that the multiplicity scaled correlation grows 
with centrality. The correlation for $\Delta \eta$ remains large, decreases 
little in the 0-30\% centrality bins, and decreases sharply beyond that.}
\label{figure9}
\end{figure*}
 
%\vspace{-0.5cm}
\begin{figure*}[ht] \centerline{\includegraphics[width=0.800\textwidth]
{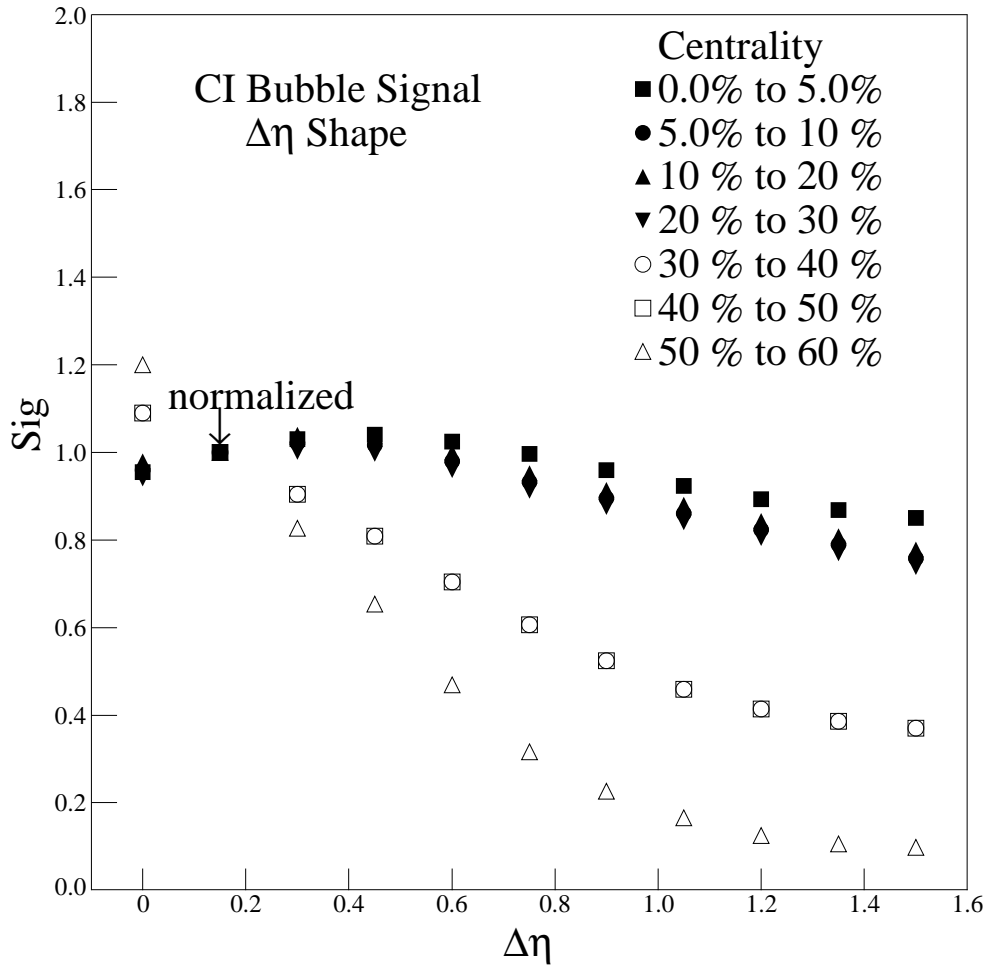}} \caption[]{The CI correlation bubble signal shape is 
normalized to one at $\Delta \eta$ equals 0.15) vs. $\Delta \eta$ for 
$10^\circ$ $< \Delta \phi <$ $20^\circ$. Seven centrality bins where there 
are bubbles present in our model are shown from 50\% to 60\% increasing to 
0\% to 5\%. The CI bubble signal $\Delta \eta$ is broad and similar (roughly
constant) as a function of $\Delta \eta$ for the 0-30\% centrality range. 
From Table and text discussion we find that bubbles in this centrality range 
contains 3-4 partons each and that it is the number of partons which determine 
the $\Delta \eta$ width. The 30-60\% centrality range show the bubble signal
$\Delta \eta$ width decreases as the number of partons decrease from 3 to 1.}
\label{figure10}
\end{figure*}

%\vspace{-0.5cm}
\begin{figure*}[ht] \centerline{\includegraphics[width=0.800\textwidth]
{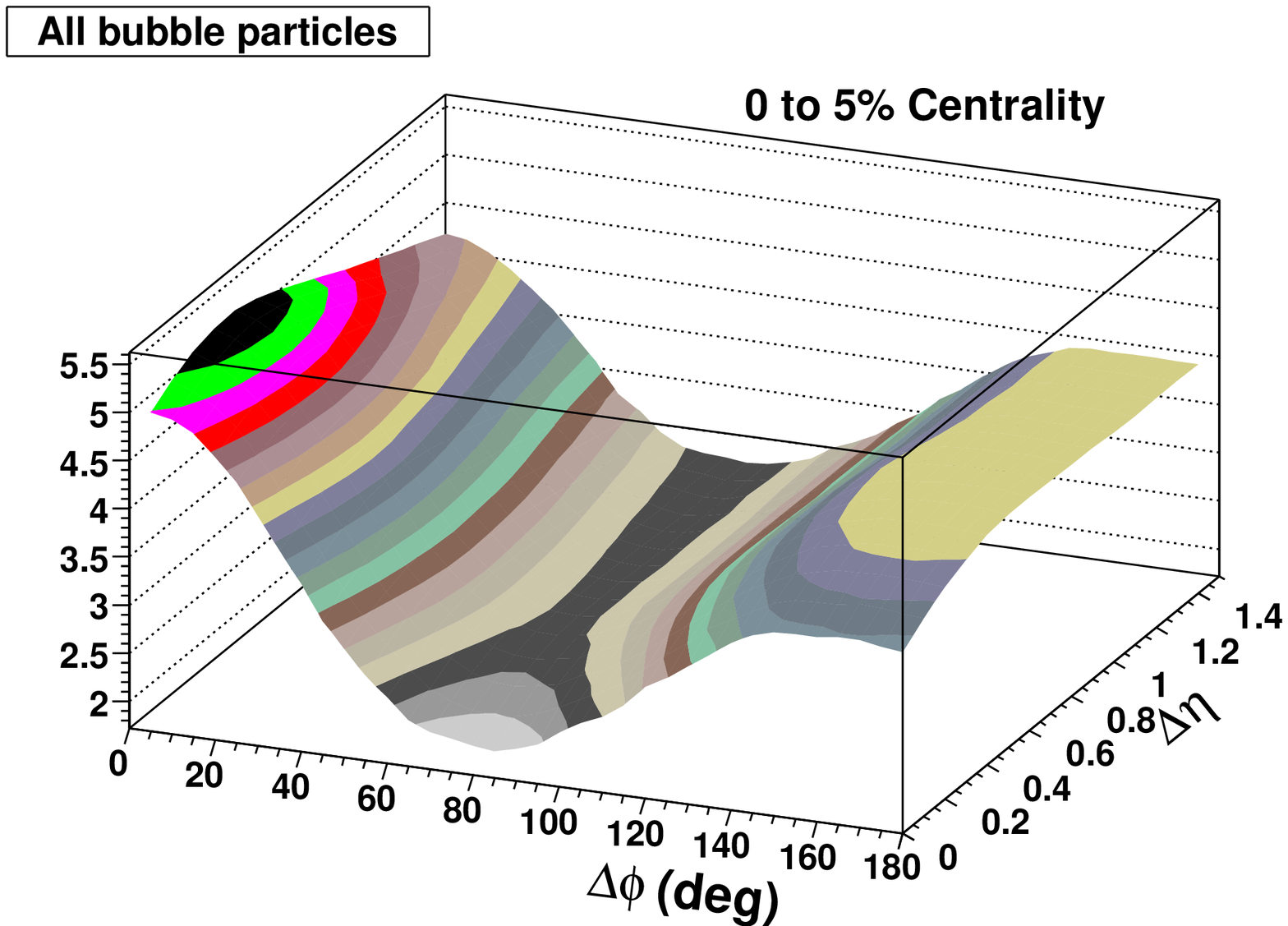}} \caption[]{``(Color online)'' The multiplicity(MULT) times 
the CI correlation for the 0-5\% centrality bin as a two dimensional 
$\Delta \phi$ vs. $\Delta \eta$ perspective plot. This is the part of the 
CI correlation that resulted from particles emitted from all the bubbles.}
\label{figure11}
\end{figure*}
 
%\vspace{-0.5cm}
\begin{figure*}[ht] \centerline{\includegraphics[width=0.800\textwidth]
{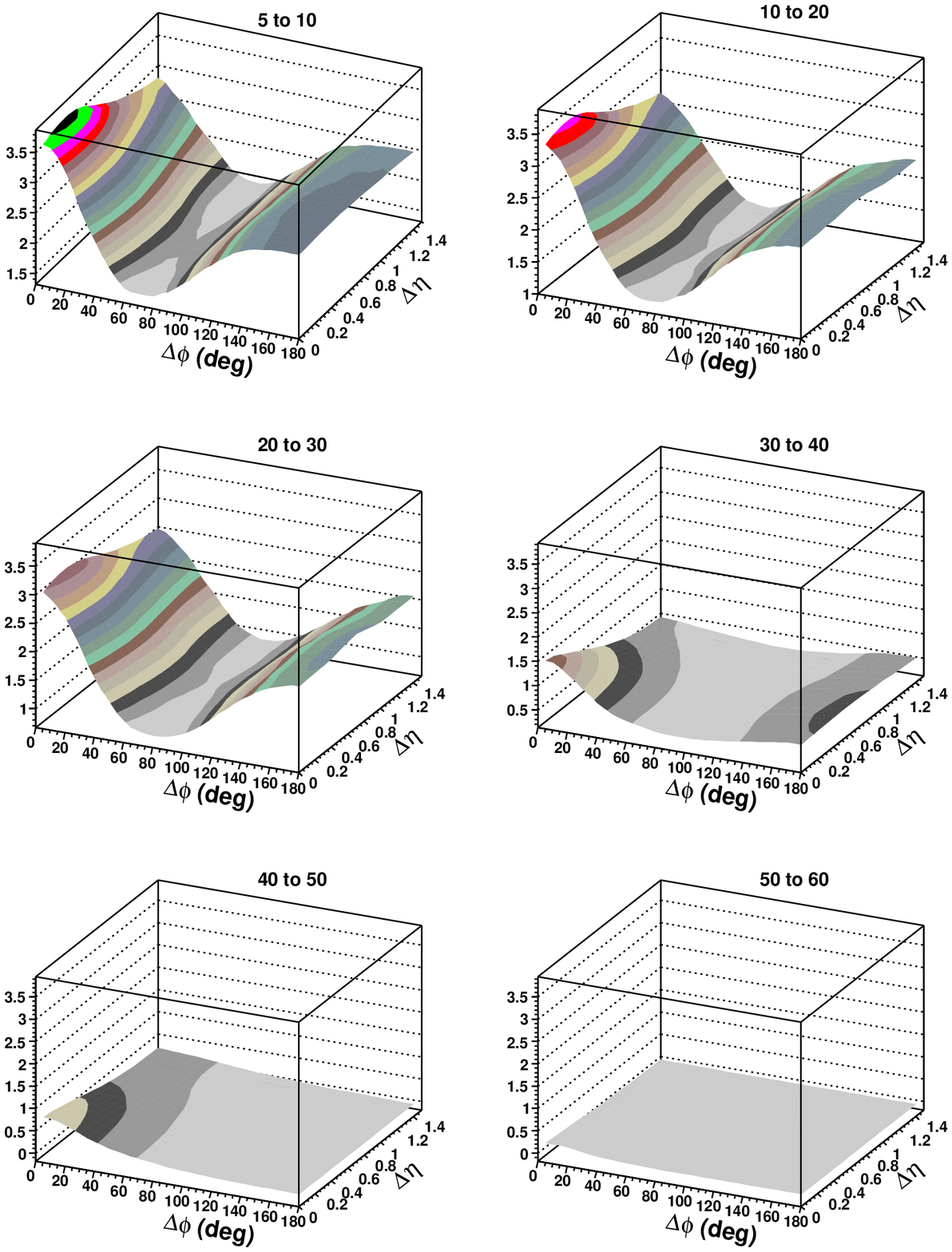}} \caption[]{``(Color online)'' The multiplicity(MULT) times 
the CI correlation for the six centralities plotted as six two dimensional 
$\Delta \phi$ vs. $\Delta \eta$ perspective plots. This is the part of the 
CI correlation that resulted from particles emitted from all the bubbles. 
We see that the multiplicity scaled CI correlation in Fig.11 and Fig.12 
from all the bubbles behaves similar to that from the CI part which
results from a single bubble as shown in Fig.8 and Fig.9 with one important
striking difference. There is a large away sige ($>$ $90^\circ$) correlation
in the 0-30\% centrality range where there is large bubble production. This 
is due to the model conserving momentum between the bubbles and is consistent
with the experimental observations.}
\label{figure12}
\end{figure*}
 
 Fig.10 shows a comparison of the bubble signal (particle from the same bubble)
shape in $\Delta \eta$ of the different centralities in which we have bubbles. 
The bubbles with the same number of partons are consistent with the same shape 
in $\Delta \eta$. Thus the longitudinal (i.e. $\Delta \eta$ distribution)
determines the number of partons in each bubble which determines the 
longitudinal shape of the bubbles. We can see that there is a maximum 
expansion in the length of $\Delta \eta$ in the most central bin (0-5\%) and 
close to this maximum is maintained through the (20-30\%) bin. However 
the length of $\Delta \eta$ considerably decreases in the (30-60\%) bins.

In Ref.\cite{PBM} the ring of bubbles played an important role in the
away side ($\Delta \phi$ $180^\circ$) correlation. Correlation between 
particles from different bubbles show an away side correlation. Fig.11
and Fig.12 show the correlation that resulted from all particles from all the 
bubbles. We plot this correlation for all centralities where there are 
bubbles present.

If we compare Fig.8 and Fig.9 (multiplicity scaled CI for particles from
the same bubble) with Fig.11 and Fig.12 (multiplicity scaled CI for particles 
from all bubbles) we note a striking difference in the away side 
($\Delta \phi > 90^\circ$) behavior of the two sets. The correlation
resulting from particles emitted by the same bubble produces an away side that
is very small (Fig.8 and Fig.9). The away side correlation produced by
particles emitted by all bubbles is large in the centrality range 0-30\% where 
bubble production is large and then essentially disappears for the more 
peripheral centralities where the bubble production becomes small to 
negligible (Fig.11 and Fig.12). This is due to the model conserving
momentum between the bubbles and is consistent with the experimental 
observations. 

Our model results in bubbles forming where energy density is highest. Thus
as we have pointed out the most central collisions have the highest energy 
density and therefore a circular ring of bubbles perpendicular to and centered
on the beam axis was able to explain the correlations 
observed in a $\sqrt{s_{NN}} =$ 200 GeV Au + Au RHIC central production 
experimental analysis\cite{PBM,centralproduction}. The PBME model with the same
bubble geometry has explained the new experimental analysis as a function of 
centrality\cite{centralitydependence}. However as we move away from central 
production bins toward peripheral bins the highest energy density 
concentrates in the region of the reaction plane, the central ring 
symmetry is broken, and the bubbles tend to be produced in the 
reaction plane region. Thus the geometry of the bubbles become coupled to 
elliptic flow. 

\section{Summary and Discussion}

In this article we summarize the assumptions made, and the reasoning that
led to the development and construction of our Parton Bubble Model\cite{PBM}
which successfully explained the charge particle pair correlations in
the central (approximately 0-10\% centrality) $\sqrt{s_{NN}}$ = 200 GeV Au +
Au data\cite{centralproduction}. The PBM was also consistent with the central
collision Au + Au HBT results. This is presented and discussed in Section 4
of Ref.\cite{PBM} and Section II of this paper. Most of this paper is 
concerned with extending our model which was a central region model to be 
able to treat the geometry of bubble production for 0-80\% collision
centralities (PBME) such as measured and analyzed in recent RHIC 
data\cite{centralitydependence}. In the PBME we included elliptic flow 
for all centralities and a jet component for centralities of 30-80\%, both 
of which become large while bubble production becomes smaller as the
centrality becomes more peripheral. We were able to extend the PBM to the 
PBME and reasonably fit the new quantitative RHIC 0-80\% $\sqrt{s_{NN}}$ = 200
Au + Au data\cite{centralitydependence} for the CI and CD correlations.
We demonstrated that the PBME had a bubble geometry that tracked the highest
energy density of the different centralities. The most central collisions 
have a symmetry about the beam axis. The region of highest energy density 
will also be symmetric about this axis and the ring of bubbles centered on 
this axis is the expected geometry nearly identical to that of the PBM.

As we move to more peripheral collisions the energy density decreases. 
The symmetry of highest energy density becomes coupled to elliptic flow 
and is defined by the reaction plane. The region of highest energy density 
becomes concentrated in the reaction plane region. Therefore 
the ring symmetry which applies to central production was broken and 
the bubbles farthest from the reaction plane disappear. The 12 bubbles 
for the most central (0-5\%) reduce to 10 (5-10\%), then to 
7 (10-20\%), and then to 5 bubbles (20-30\%). Moving to 30-40\%, 
40-50\%, and 50-60\% the bubbles are in the reaction plane region with the 
probability per event of making a bubble being 1.5, 0.6, 0.3 (see Table I).

Both the PBM and the PBME treat the fireball in a blast wave model. 
Bubbles are formed on the surface of the blast wave fireball 
which emit the final state particles at kinetic freezeout. We 
achieve reasonable fits to the quantitative STAR experimental 
analysis of the Charge Independent (CI) and the Charge Dependent (CD) 
correlations within a few percent of the total 
correlations\cite{centralitydependence}. These correlations 
considered particles in the $p_t$ range 0.8 to 4.0 GeV/c. The model is also 
consistent with the HBT results in this $p_t$ range. The CI 
correlation displays the average structure of the correlated emitting
sources at kinetic freezeout. The CD correlation has the initial emission 
correlation of the opposite sign charge pairs of particles emitted from the
same space-time region at the time of hadronization. This initial correlation 
should remain consistent with the CD correlation of Pythia jets if there is no
further interactions with other particles after hadronization. The analysis 
above demonstrated that at kinetic freezeout the CD correlation in the above 
$p_t$ range was consistent with Pythia jets at all centralities 
(see Section VI). Therefore one concludes that hadronization and kinetic 
freezeout both occur at or very near the surface of the fireball at all 
centralities. Thus the expanding fireball is dense and opaque for most 
centralities at kinetic freezeout. Of course the most peripheral bins due to 
the low material content allow small path length to the surface. Both the PBM 
and its extended present version PBME are surface emission models and thus are 
consistent with this striking experimental feature.

It is of interest to note that the parton bubble model central collision 
analysis\cite{PBM} has recently been pointed out in \cite{glasma} as having
features in common with glasma flux tubes which evolve from initial color 
glass condensates. This supports the hypothesis that the bubble substructure 
can be considered a QGP signal and may serve as a key in investigating both
QGP and glasma effects.

The persistence of the production of similar surface bubbles at kinetic 
freezeout in numbers which decrease as the highest energy density decreases 
as the centrality is decreased (going toward the peripheral bins); and their 
general characteristic of being produced 
in the region of highest energy density implies the following:

1) The bubbles represent a significant substructure of gluonic hot spots 
formed on the surface of a dense opaque fireball at kinetic freezeout. The 
number of bubbles formed and the energy content of each of these
substructures is a function of the energy density and its extent in space.

2) Their characteristics, persistence, behavior as a function of centrality,
and the PBME reasonably quantitative fits to the CI and CD data at all 
centralities provide substantial evidence that the bubbles are the final state
products of QGP production. If sufficiently convincing QGP signatures can be 
extracted from these bubbles then one could eventually provide substantial 
evidence for a QGP. We will be investigating this in future work. The bubbles 
may contain relevant information on other topics of interest (e.g. glasma)
which we will also investigate. In our future program we plan to utilize 
the anticipated forthcoming availability of Time of Flight particle 
identification data at STAR to further study charged particles 
correlations with identified particles. In the second paragraph from the 
end of Section 1.2 of Ref.\cite{PBM} we speculated on the possibility 
of applying these model ideas (suitably modified) to LHC data when 
it becomes available.
  
\section{acknowledgements}

The authors thank William Love for valuable discussion and assistance in 
production of figures. This research was supported by the U.S. Department of 
Energy under Contract No. DE-AC02-98CH10886 and the City College of New York 
Physics Department of the City University of New York.

\end{document}